\documentclass[11pt,a4paper]{article}
\usepackage[numbers,sort&compress]{natbib}
\usepackage{amsmath,amssymb,amsthm}
\usepackage[margin=1in]{geometry}
\usepackage[english]{babel}
\usepackage{hyperref}
\usepackage{graphicx}
\usepackage{physics}
\usepackage{bm}
\usepackage{caption}
\usepackage{subcaption}
\usepackage{color}
\usepackage{fancybox}
\usepackage{amsmath}
\usepackage{amsfonts}
\usepackage{cancel}
\usepackage{mathrsfs}
\usepackage{multirow}
\usepackage{pstricks}
\usepackage{multicol}
\usepackage{bm}
\usepackage{tikz}
\usepackage{tikz-feynman}
\usepackage{feynmp-auto}
\usepackage{amssymb}
\usepackage{float}
\usepackage{xcolor}
\usepackage{xfrac}
\usepackage{chemfig}
\usepackage{eso-pic}
\usepackage[numbers]{natbib}
\usepackage[T1]{fontenc}
\usepackage{titlesec}
\usepackage{authblk}
\usepackage{enumitem}
\titleformat{\section}
{\bfseries}{\thesection.}{1em}{\normalfont\bfseries}
\usetikzlibrary{positioning, arrows.meta}
\titleformat{\subsection}
{\bfseries}{\thesubsection}{1em}{\small\bfseries}
\usetikzlibrary{positioning, arrows.meta}
\hypersetup{
	colorlinks=true,       
	linkcolor=red,       
	citecolor=green,     
	filecolor=magenta,     
	urlcolor=cyan          
}
\theoremstyle{definition}

\numberwithin{equation}{section}

\title{\large\textbf{Ward-Takahashi Identity in Denominator Regularization at One Loop}}
\author[1,2]{Mickaya A. Razanaparany\footnote{\small Email: \href{mailto:mickaya@aims.ac.za}{\texttt{mickaya@aims.ac.za}}}}
\affil[1]{\small\textit{Physique des Hautes Energies, PHE, Université d'Antananarivo, Madagascar}}
\affil[2]{\small\textit{African Institute for Mathematical Sciences, AIMS, South Africa}}
\date{}
\begin{document}
\maketitle
\begin{abstract}
	Explicit analytic expressions for the electron self-energy and the vertex correction in quantum electrodynamics are derived at one loop using the recently proposed regularization scheme known as denominator regularization, assisted by its correspondence with dimensional regularization to determine the coefficient functions, which are a specific ingredient of this approach. We then show that the regularized amplitudes satisfy the Ward-Takahashi identity, thereby ensuring that gauge symmetry is preserved after regularization.
\end{abstract}
\section{Introduction}
Loop diagrams derived from Feynman rules in quantum field theory \texttt{(QFT)} often produce divergent integrals. These divergences are treated using regularization, which ensures that physical quantities remain finite when the regulator is removed. Several regularization schemes exist, each with its own advantages and limitations. A crucial requirement is that the chosen scheme preserves the symmetries of the theory \cite{peskin, grozin, tarrach, schwartz}.

Dimensional regularization \texttt{(DIM REG)} is one of the most commonly used schemes in \texttt{QFT}. It controls divergent loop integrals by analytically continuing the number of spacetime dimensions from $4$ to $d=4-2\epsilon$ \cite{thooft}. By slightly reducing the number of dimensions, the momentum integrals grow more slowly at high energies, which makes the previously divergent integrals finite. The divergences then appear as poles in $\frac{1}{\epsilon}$ when the result is expanded around $\epsilon=0$, and these poles are removed through renormalization \cite{peskin, tarrach, schwartz}.

Denominator regularization \texttt{(DEN REG)} is a recently proposed scheme in which, after combining terms with Feynman parameters, the overall power of the denominator in a loop amplitude is analytically continued from $n$ to $n+\epsilon$, while the number of spacetime dimensions is fixed. A specific ingredient of this scheme is the coefficient function $f_{(n,p)}(\epsilon)$, which smoothly approaches $1$ as $\epsilon \to 0$, where $n$ is the original power of the denominator and $p$ is the superficial degree of divergence \cite{will_and_plessis, will}. \texttt{DEN REG} is compatible with the minimal subtraction (\texttt{MS}) renormalization scheme, and some of its desirable properties and advantages are discussed in \cite{will, will_and_plessis, will_and_plessis_2, wah2023, more}.

Our objective is to perform a consistency check of the Ward-Takahashi identity (\texttt{WTI}) using \texttt{DEN REG}. The \texttt{WTI} is useful for demonstrating that gauge symmetry is preserved after regularization \cite{peskin, grozin, tarrach, schwartz, wardre, takahashibook}. In this work, we explore the application of \texttt{DEN REG} to this problem, showing how this scheme can be employed to regularize the electron self-energy and vertex correction at one-loop level, leading to the verification of the \texttt{WTI} at next-to-leading order (\texttt{NLO}), under the on-shell condition.

The structure of the rest of this paper is as follows. In Section \ref{correspondance_sec}, we present the correspondence between \texttt{DIM REG} and \texttt{DEN REG}. Section \ref{loopDENREGcalc766} focuses on the one loop calculation of the electron self-energy and vertex correction using the \texttt{DEN REG} scheme. In Section \ref{wtiiiiivery}, we provide a straightforward verification of the \texttt{WTI}. We conclude in Section \ref{cclDNWTI}. Important integrals in \texttt{DEN REG}, as well as details of the computation of the numerator of the vertex correction in $4$ dimensions, are presented in Appendices \ref{appendix_1} and \ref{appendix_2}, respectively.
\section{Correspondence Between Dimensional and Denominator Regularization}\label{correspondance_sec}
In \cite{will_and_plessis_2}, a natural setting and manipulation are employed to move from \texttt{DIM REG} to \texttt{DEN REG}. The manipulation proceeds as follows: one begins with the idea of \texttt{DIM REG}, in which the number of spacetime dimensions is analytically continued, setting $d = 4-2\epsilon$, and then obtains the corresponding result in \texttt{DEN REG}, where the spacetime dimension is fixed and the power of the single denominator is analytically continued to $n+\epsilon$ with a coefficient function that depends on $\epsilon$ and approaches $1$ as $\epsilon \rightarrow 0$. The correspondences between \texttt{DIM REG} and \texttt{DEN REG} are as follows:
\begin{equation}\label{cor_1}
	\begin{tikzpicture}[baseline]
		\node (A) at (0,0) {$\displaystyle \mu^{4-d}\int\frac{d^d k}{(2\pi)^d} \frac{1}{(k^2+\Delta^2)^2}$};
		\node (B) at (8,0) {$\displaystyle (4\pi)^\epsilon \Gamma(2+\epsilon) \mu^{2\epsilon} \int \frac{d^4 k}{(2\pi)^4} \frac{1}{(k^2+\Delta^2)^{2+\epsilon}},$};
		
		\draw[->, thick] ([yshift=0.3em]A.east) -- ([yshift=0.3em]B.west);
		\node at (3.4,0.3) {\small \texttt{DEN REG}};
		
		\draw[->, thick] ([yshift=-0.3em]B.west) -- ([yshift=-0.3em]A.east);
		\node at (3.4,-0.3) {\small \texttt{DIM REG}};
	\end{tikzpicture}
\end{equation}
\begin{equation}\label{cor_2}
	\begin{tikzpicture}[baseline]
		\node (A) at (0,0) {$\displaystyle \mu^{4-d}\int\frac{d^d k}{(2\pi)^d} \frac{k^2}{(k^2+\Delta^2)^2}$};
		\node (B) at (8,0) {$\displaystyle (4\pi)^\epsilon\Gamma(1+\epsilon)\mu^{2\epsilon}\int\frac{d^4k}{(2\pi)^4}\frac{k^2-\epsilon\Delta^2}{(k^2+\Delta^2)^{2+\epsilon}},$};
		
		\draw[->, thick] ([yshift=0.3em]A.east) -- ([yshift=0.3em]B.west);
		\node at (3.4,0.3) {\small \texttt{DEN REG}};
		
		\draw[->, thick] ([yshift=-0.3em]B.west) -- ([yshift=-0.3em]A.east);
		\node at (3.4,-0.3) {\small \texttt{DIM REG}};
	\end{tikzpicture}
\end{equation}
\begin{equation}\label{cor_3}
	\begin{tikzpicture}[baseline]
		\node (A) at (0.1,0) {$\displaystyle \mu^{4-d}\int \frac{d^d k}{(2\pi)^d} \frac{1}{(k^2 + \Delta^2)^3}$};
		\node (B) at (8,0) {$\displaystyle \frac{(4\pi)^\epsilon}{2}\Gamma(3+\epsilon)\mu^{2\epsilon}\int \frac{d^4 k}{(2\pi)^4} \frac{1}{(k^2 + \Delta^2)^{3+\epsilon}},$};
		
		\draw[->, thick] ([yshift=0.3em]A.east) -- ([yshift=0.3em]B.west);
		\node at (3.5,0.3) {\small \texttt{DEN REG}};
		
		\draw[->, thick] ([yshift=-0.3em]B.west) -- ([yshift=-0.3em]A.east);
		\node at (3.5,-0.3) {\small \texttt{DIM REG}};
	\end{tikzpicture}
\end{equation}
\begin{equation}\label{cor_4}
	\begin{tikzpicture}[baseline]
		\node (A) at (0,0) {$\displaystyle \displaystyle\mu^{4-d}\int \frac{d^dk}{(2\pi)^d}\frac{k^2}{(k^2+\Delta^2)^3}$};
		\node (B) at (8,0) {$\displaystyle (4\pi)^\epsilon\Gamma(2+\epsilon)\mu^{2\epsilon}\int \frac{d^4k}{(2\pi)^4}\frac{k^2-\frac{1}{2}\epsilon\Delta^2}{(k^2+\Delta^2)^{3+\epsilon}}.$};
		
		\draw[->, thick] ([yshift=0.3em]A.east) -- ([yshift=0.3em]B.west);
		\node at (3.4,0.3) {\small \texttt{DEN REG}};
		
		\draw[->, thick] ([yshift=-0.3em]B.west) -- ([yshift=-0.3em]A.east);
		\node at (3.4,-0.3) {\small \texttt{DIM REG}};
	\end{tikzpicture}
\end{equation}
Both in \texttt{DIM REG} and \texttt{DEN REG}, Eqs.~\eqref{cor_1}--\eqref{cor_4} have exactly the same left- and right-hand sides when we set $d = 4$, that is, $\epsilon = 0$. Note that these integrals appear after combining the denominators using Feynman parameters and performing a Wick rotation. Here, $\Gamma$ denotes the Gamma function, one of its most useful properties is $\Gamma(1+\epsilon)=\epsilon\Gamma(\epsilon)$.

We can check the self-consistency of these correspondences by computing each side straightforwardly. Let $L_i(\mu,\Delta,\epsilon)$ and $R_i(\mu,\Delta,\epsilon)$, for $i = 1,2,3,4$, denote the left-hand side and right-hand side, respectively, of Eqs.~\eqref{cor_1}--\eqref{cor_4}. We then obtain the following results:
\begin{align}
	L_1(\mu, \Delta, \epsilon)&=R_1(\mu, \Delta, \epsilon)=(4\pi)^{\epsilon-2}\left(\frac{\mu^2}{\Delta^2}\right)^\epsilon\Gamma(\epsilon),\label{consistency_1}\\
	L_2(\mu, \Delta, \epsilon)&=R_2(\mu, \Delta, \epsilon)= (4\pi)^{\epsilon-2}\left(\frac{\mu^2}{\Delta^2}\right)^\epsilon\Delta^2(2-\epsilon)\Gamma(\epsilon-1),\label{consistency_2}\\
	L_3(\mu, \Delta, \epsilon)&=R_3(\mu, \Delta, \epsilon)=(4\pi)^{\epsilon-2}\left(\frac{\mu^2}{\Delta^2}\right)^\epsilon\frac{1}{2\Delta^2}\Gamma(1+\epsilon),\label{consistency_3}\\
	L_4(\mu, \Delta, \epsilon)&=R_4(\mu, \Delta, \epsilon)=(4\pi)^{\epsilon-2}\left(\frac{\mu^2}{\Delta^2}\right)^\epsilon\frac{1}{2}(2-\epsilon)\Gamma(\epsilon).\label{consistency_4}
\end{align}
These are easy to verify using a change of variables and based on the integrals in Eqs.~\eqref{integral_1}--\eqref{integral_4}. All follow an analogous procedure.
\section{One Loop Denominator Regularization in Quantum Electrodynamics}\label{loopDENREGcalc766}
This section contains the detailed calculations of \texttt{DEN REG} in the context of the one loop electron self-energy and vertex correction for \texttt{QED}. We highlight the \texttt{DEN REG} procedure and demonstrate its versatility, which is particularly advantageous for the vertex correction, as only the computation of the numerator in four dimensions is required (see Appendix \ref{appendix_2}), making it less subtle and technically simpler than in $d$ dimensions.
\begin{figure}
	\centering
	\begin{subfigure}{0.45\textwidth}
		\centering
		\begin{tikzpicture}
			\begin{feynman}
				\vertex (a);
				\vertex [right=of a] (b);
				\vertex [right=of b] (c);
				\vertex [right=of c] (d);
				\diagram*{
					(a) -- [fermion, edge label'= $p$] (b) -- [photon, half left, looseness=1.5, momentum=$p-k$] (c) -- [fermion, edge label'= $p$] (d),
					(b) -- [fermion, edge label'= $k$] (c),
				};
			\end{feynman}
		\end{tikzpicture}
		\caption{}
		\label{fig:one_loopFIGY_a}
	\end{subfigure}\hfill
	\begin{subfigure}{0.45\textwidth}
		\centering
		\begin{tikzpicture}
			\begin{feynman}
				\tikzfeynmanset{every vertex={dot, minimum size=0pt}}
				\vertex (a) at (0,0);
				\vertex [right=2cm of a] (b);
				\vertex [above right=0.7cm and 1cm of b] (c);
				\vertex [below right=0.7cm and 1cm of b] (d);
				\vertex [right=1.5cm of c] (e) at (2.3,1.3);
				\vertex [right=1.5cm of d] (f) at (2.3,-1.3);
				\diagram* {
					(a) -- [boson, decoration={snake}, momentum=\(q\)] (b),
					(b) -- [fermion, edge label=\(p^\prime+k\)] (c) -- [fermion, edge label=\(p'\)] (e),
					(f) -- [fermion, edge label=\(p\)] (d) -- [fermion, edge label=\(p+k\)] (b),
					(c) -- [boson, decoration={snake}, momentum=\(k\)] (d)
				};
			\end{feynman}
		\end{tikzpicture}
		\caption{}
		\label{fig:one_loopFIGY_b}
	\end{subfigure}
	\caption{\texttt{NLO} corrections in \texttt{QED}: (a) self-energy loop modifying the electron propagator, (b) vertex loop modifying the electron-photon coupling.}
	\label{fig:one_loopFIGY}
\end{figure}
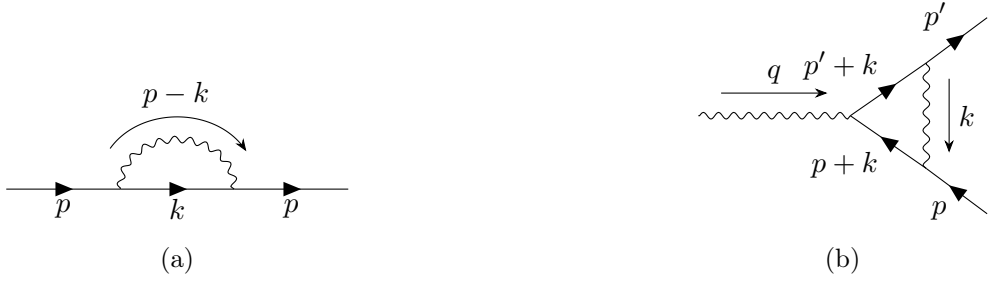
\subsection{Electron Self-Energy}
The one loop correction to the electron self-energy in \texttt{QED} is represented by the Feynman diagram shown in Fig.~\ref{fig:one_loopFIGY_a}. According to the Feynman rules, the corresponding expression is given by
\begin{equation}
	-i\sum_{\text{1-Loop}}(\cancel{p})=-e^2\int\frac{d^4k}{(2\pi)^4}\frac{\gamma^\mu(\cancel{k}+m)\gamma_\mu}{[k^2-m^2+i\epsilon][(p-k)^2+i\epsilon]}.\label{elctAfterRules}
\end{equation}
In order to combine the denominators of the propagators, we use Feynman parametrization and denote the shifted momentum by $\ell := k-xp$, we find:
\begin{equation}\label{linear_vanished}
	\begin{aligned}
			-i\sum_{\text{1-Loop}}(\cancel{p})&=-e^2\int_{0}^{1}dx\int\frac{d^4\ell}{(2\pi)^4}\frac{\gamma^\mu(\cancel{\ell}+x\cancel{p})\gamma_\mu+m\gamma^\mu\gamma_\mu}{(\ell^2-\Delta^2)^2}\\
			&=-e^2\int_{0}^{1}dx\int\frac{d^4\ell}{(2\pi)^4}\frac{x\gamma^\mu\cancel{p}\gamma_\mu+m\gamma^\mu\gamma_\mu}{(\ell^2-\Delta^2)^2},
	\end{aligned}
\end{equation}
where $\Delta^2\equiv(1-x)m^2-(1-x)xp^2$. In the second line of Eq.~\eqref{linear_vanished}, the term linear in $\ell$ is omitted, since it does not contribute owing to its odd parity. Employing $\gamma^\mu \gamma_\mu = 4$ and $\gamma^\mu \cancel{p} \gamma_\mu = -2 \cancel{p}$, the \texttt{DEN REG} prescription allows us to write
\begin{equation}
	\begin{tikzpicture}[baseline]
		\node (A) at (0,0) {$\displaystyle -i\sum_{\text{1-Loop}}(\cancel{p})$}; 
		\node (B) at (7,0) {$\displaystyle -2e^2(-\mu^2)^\epsilon f_{(2,0)}(\epsilon)\int_{0}^{1}dx\int \frac{d^4\ell}{(2\pi)^4}\frac{-x\cancel{p}+2m}{(\ell^2-\Delta^2)^{2+\epsilon}}.$};
		
		\draw[->, thick] (A) -- (B);
		\node at (1.9,0.2) {\small\texttt{DEN REG}}; 
	\end{tikzpicture}
\end{equation}
Here, a dimensionful scale $\mu$ is introduced to maintain the dimensionality of $\sum\limits_{\text{1-Loop}}(\cancel{p})$, and the minus sign accompanying $\mu$ cancels the factor $(-1)^{-\epsilon}$ arising from the \texttt{DEN REG} procedure after the Wick rotation. We now perform the Wick rotation, obtaining:
\begin{align}
	-i\sum_{\text{1-Loop}}(\cancel{p})&=-2ie^2(-\mu^2)^\epsilon f_{(2,0)}(\epsilon)\int_{0}^{1}dx\int \frac{d^4\ell_E}{(2\pi)^4}\frac{-x\cancel{p}+2m}{(-1)^{2+\epsilon}(\ell^2_E+\Delta^2)^{2+\epsilon}}\nonumber\\
	&=\frac{4e^2}{(4\pi)^4}\int_{0}^{1}dx(-x\cancel{p}+2m)f_{(2,0)}(\epsilon)\int_{0}^{\infty}d\ell_E\frac{\mu^{2\epsilon}\ell_E^3}{(\ell_E^2+\Delta^2)^{2+\epsilon}}.\label{afete_Wick55}
\end{align}
From the correspondence given in Eq.~\eqref{cor_1}, we can obtain a coefficient function
\begin{equation}\label{coeff_f1}
	f_{(2,0)}(\epsilon)=(4\pi)^\epsilon\Gamma(2+\epsilon).
\end{equation}
We utilize the integral in Appendix \ref{appendix_1}, substituting Eqs.~\eqref{integral_1} and \eqref{coeff_f1} into Eq.~\eqref{afete_Wick55}, and carefully using the expansion of the Gamma function $\Gamma(\epsilon)$ near $\epsilon = 0$, we obtain the following analytical result for the electron self-energy
\begin{equation}
	\sum_{\text{1-Loop}} (\cancel{p})=\frac{2e^2}{(4\pi)^2}\int_{0}^{1}dx(-x\cancel{p}+2m)\left[\frac{1}{\epsilon}-\gamma_E+\ln(4\pi)+\ln\left(\frac{\mu^2}{\Delta^2}\right)\right]+\mathcal{O}(\epsilon),
\end{equation}
with $\gamma_E$ denoting the Euler-Mascheroni constant. These expressions provide the correct form of the divergent parts, the $\frac{1}{\epsilon}$ pole arising from \texttt{DEN REG} captures the logarithmic \texttt{UV} divergence of the integral.
\subsection{Vertex Corrections}
The Feynman diagram shown in Fig.~\ref{fig:one_loopFIGY_b} corresponds to the vertex correction in \texttt{QED} at the one loop level. Using the Feynman rules, the contribution of this diagram can be evaluated as follows:
\begin{align}
	ie\Gamma_{\text{1-Loop}}^\mu(p^\prime, p)&=e^3\int \frac{d^4k}{(2\pi)^4}\frac{\gamma^\nu(\cancel{p}^\prime+\cancel{k}+m)\gamma^\mu(\cancel{p}+\cancel{k}+m)\gamma_\nu}{[(p^\prime+k)^2-m^2+i\varepsilon][(p+k)^2-m^2+i\varepsilon][k^2+i\varepsilon]}\nonumber\\
	&:=e^3\int \frac{d^4k}{(2\pi)^4}\frac{\mathcal{N}^\mu}{\mathcal{D}},\label{loopGAMM}
\end{align}
where
\begin{equation}\label{num_VERTEX_Calc}
	\mathcal{N}^\mu=\gamma^\nu(\cancel{p}^\prime+\cancel{k}+m)\gamma^\mu(\cancel{p}+\cancel{k}+m)\gamma_\nu,
\end{equation}
\begin{equation}
	\mathcal{D}=[(p^\prime+k)^2-m^2+i\varepsilon][(p+k)^2-m^2+i\varepsilon][k^2+i\varepsilon].
\end{equation}
We use Feynman parametrization to combine the denominators, and we have:
\begin{equation*}
	\frac{1}{\mathcal{D}}=\int_0^1 \int_0^1 \int_0^1 dxdydz\delta(x+y+z-1)\frac{2}{\left[x((p+k)^2-m^2)+y((p^\prime+k)^2-m^2)+zk^2\right]^3}.
\end{equation*}
We can compute the contents inside the square brackets as follows:
\begin{equation*}
	[\Box]=x((p+k)^2-m^2)+y((p^\prime+k)^2-m^2)+zk^2:=\ell^2-\Delta^2,
\end{equation*}
where $\ell:=k+xp+yp^\prime$ and $\Delta^2:=(1-z)^2m^2-xyq^2$, this follows from the on-shell condition for the electron momenta, $p^{\prime^2}=p^2=m^2$. Then, the loop integral in Eq.~\eqref{loopGAMM} can be written as:
\begin{equation}\label{gamVERTf089}
	\Gamma_{\text{1-Loop}}^\mu(p^\prime, p)=-2ie^2\int_0^1 d\mathbb{X}_3\int \frac{d^4\ell}{(2\pi)^4}\frac{\mathcal{N}^\mu}{(\ell^2-\Delta^2)^3},
\end{equation}
where, to avoid writing out lengthy expressions, we have denoted the triple integrals as
\begin{equation*}
	\int_0^1 d\mathbb{X}_3:=\int_0^1\int_0^1\int_0^1 dxdydz\delta(x+y+z-1).
\end{equation*}
Now, in terms of \texttt{DEN REG}, the number of spacetime dimensions is fixed at 4. The evaluation of the numerator in 4 dimensions is given in detail in Appendix~\ref{appendix_2}. Our expression for $\Gamma_{\text{1-Loop}}^\mu(p^\prime, p)$ in Eq.~\eqref{gamVERTf089} transforms as follows:
\begin{equation}
	\begin{tikzpicture}[baseline]
		\node (A) at (0,0) {$\displaystyle \Gamma_{\text{1-Loop}}^\mu(p^\prime, p)$}; 
		\node (B) at (5.3,0) {$\displaystyle F_1(q^2)\gamma^\mu+F_2(q^2)\frac{i\sigma^{\mu\nu}q_\nu}{2m},$};
		
		\draw[->, thick] (A) -- (B);
		\node at (2.1,0.2) {\small\texttt{DEN REG}}; 
	\end{tikzpicture}
\end{equation}
where the corresponding form factors are given by:
\begin{align*}
	F_1(q^2)&=-2ie^2(-\mu^2)^{\epsilon}\int_0^1d\mathbb{X}_3\int\frac{d^4\ell}{(2\pi)^4}\frac{f_{(3,1)}(\epsilon)\ell^2-2g_{(3,1)}(\epsilon)\Delta^2+2z(2m^2-q^2)f_{(3,0)}(\epsilon)}{(\ell^2-\Delta^2)^{3+\epsilon}},\\
	F_2(q^2)&=+2ie^2(-\mu^2)^{\epsilon}\int_0^1 d\mathbb{X}_3\int\frac{d^4\ell}{(2\pi)^4}\frac{4f_{(3,0)}(\epsilon)(1-z)zm^2}{(\ell^2-\Delta^2)^{3+\epsilon}}.
\end{align*}
The $-(-1)^\epsilon$ cancels the $-(-1)^{-\epsilon}$ introduced by \texttt{DEN REG} after Wick rotation. We can rewrite those two expressions as the following form:
\begin{equation*}
	\begin{aligned}
		F_1(q^2)&=-2ie^2(-1)^\epsilon\int_0^1d\mathbb{X}_3\bigg\{f_{(3,1)}(\epsilon)\mathbb{I}_2(\epsilon)+\big[-2g_{(3,1)}(\epsilon)\Delta^2+2z(2m^2-q^2)f_{(3,0)}(\epsilon)\big]\mathbb{I}_1(\epsilon)\bigg\},\\
		F_2(q^2)&=+2ie^2(-1)^\epsilon\int_0^1 d\mathbb{X}_3\bigg\{4f_{(3,0)}(\epsilon)(1-z)zm^2\mathbb{I}_1(\epsilon)\bigg\}.
	\end{aligned}
\end{equation*}
where the two integrals $\mathbb{I}_1(\epsilon)$ and $\mathbb{I}_2(\epsilon)$ are defined by:
\begin{equation}
	\mathbb{I}_1(\epsilon):=\int\frac{d^4\ell}{(2\pi)^4}\frac{\mu^{2\epsilon}}{(\ell^2-\Delta^2)^{3+\epsilon}},\quad\text{and}\quad
	\mathbb{I}_2(\epsilon):=\int\frac{d^4\ell}{(2\pi)^4}\frac{\mu^{2\epsilon}\ell^2}{(\ell^2-\Delta^2)^{3+\epsilon}}.
\end{equation}
Our next task is to evaluate the integrals $\mathbb{I}_1(\epsilon)$ and $\mathbb{I}_2(\epsilon)$:
\begin{align}
	\mathbb{I}_1(\epsilon)\longrightarrow&-i(-1)^{-\epsilon}\int\frac{d^4\ell_E}{(2\pi)^4}\frac{\mu^{2\epsilon}}{(\ell_E^2+\Delta^2)^{3+\epsilon}}\qquad\texttt{(Wick rotation)}\nonumber\\
	=&-i(-1)^{-\epsilon}\frac{2\pi^2}{(2\pi)^4}\int_{0}^{\infty}d\ell_E\frac{\mu^{2\epsilon}\ell_E^3}{(\ell_E^2+\Delta^2)^{3+\epsilon}}\qquad\texttt{\eqref{integral_3}}\nonumber\\
	=&-i(-1)^{-\epsilon}\frac{1}{(4\pi)^2}\left(\frac{\mu^2}{\Delta^2}\right)^\epsilon\frac{1}{\Delta^2}\frac{1}{(1+\epsilon)(2+\epsilon)},\label{Iun}
\end{align}
and
\begin{align}
	\mathbb{I}_2(\epsilon)\longrightarrow&+i(-1)^{-\epsilon}\int\frac{d^4\ell_E}{(2\pi)^4}\frac{\mu^{2\epsilon}\ell_E^2}{(\ell_E^2+\Delta^2)^{3+\epsilon}}\qquad\texttt{(Wick rotation)}\nonumber\\
	=&+i(-1)^{-\epsilon}\frac{2\pi^2}{(2\pi)^4}\int_{0}^{\infty}d\ell_E\frac{\mu^{2\epsilon}\ell_E^5}{(\ell_E^2+\Delta^2)^{3+\epsilon}}\qquad\texttt{\eqref{integral_4}}\nonumber\\
	=&+i(-1)^{-\epsilon}\frac{2}{(4\pi)^2}\left(\frac{\mu^2}{\Delta^2}\right)^\epsilon\frac{1}{\epsilon}\frac{1}{(1+\epsilon)(2+\epsilon)}\label{Ideux}.
\end{align}
From the correspondences in Eqs.~\eqref{cor_3} and \eqref{cor_4}, the coefficient functions of interest are:
\begin{equation}\label{coeff_values}
	f_{(3,0)}(\epsilon) = \frac{(4\pi)^\epsilon}{2}\Gamma(3+\epsilon),\quad
	f_{(3,1)}(\epsilon) = (4\pi)^\epsilon\Gamma(2+\epsilon)\quad\text{and}\quad
	g_{(3,1)}(\epsilon) = -\frac{(4\pi)^\epsilon}{2}\epsilon\Gamma(3+\epsilon).
\end{equation}
Note that $g_{(3,1)}(\epsilon)$ is an exceptional coefficient function that goes to zero as $\epsilon\to 0$. Substituting Eqs.~\eqref{Iun}, \eqref{Ideux} and \eqref{coeff_values} into the two form factors and simplifying carefully, we arrive at the final analytical results for the one-loop vertex correction in \texttt{QED}:
\begin{equation}\label{resultDENREG_VeRt}
	\Gamma_{\text{1-Loop}}^\mu(p^\prime, p)=F_1(q^2)\gamma^\mu+F_2(q^2)\frac{i\sigma^{\mu\nu}q_\nu}{2m},
\end{equation}
with the form factors:
\begin{align}
	F_1(q^2)&=\frac{2e^2}{(4\pi)^2}\int_{0}^{1}d\mathbb{X}_3\left[\frac{1}{\epsilon}-\gamma_E+\ln(4\pi)+\ln\left(\frac{\mu^2}{\Delta^2}\right)-\frac{z(2m^2-q^2)}{\Delta^2}\right]+\mathcal{O}(\epsilon)\\
	F_2(q^2)&=\frac{2e^2}{(4\pi)^2}\int_{0}^{1}d\mathbb{X}_3\left[\frac{2(1-z)zm^2}{\Delta^2}\right]+\mathcal{O}(\epsilon).
\end{align}
The form factors $F_1(q^2)$ and $F_2(q^2)$ encode the effects of the one-loop \texttt{QED} vertex correction. The Dirac form factor $F_1(q^2)$ contains both the \texttt{UV} divergence, represented by the $\frac{1}{\epsilon}$ pole in \texttt{DEN REG}, and finite momentum-dependent contributions, reflecting charge renormalization. In contrast, the Pauli form factor $F_2(q^2)$ is finite in the limit $\epsilon\to 0$ and depends on the Feynman parameters $x$, $y$ and $z$ through $\Delta^2 = (1-z)^2 m^2-xy q^2$. In the static limit $q^2 \to 0$, $F_2(0)$ directly yields the anomalous magnetic moment of the electron, giving the well-known one-loop result $F_2(0) = \frac{\alpha}{2\pi}$ \citep{kush_foley, schwinger}.

\section{Consistency with the Ward-Takahashi Identity}\label{wtiiiiivery}
In this section, we demonstrate that gauge invariance is preserved under the \texttt{DEN REG} scheme. To this end, we begin with the Feynman diagrammatic representation of the \texttt{WTI}, shown in Fig.~\ref{fig:WTI_diagram86486}, and subsequently verify it using our previous results obtained within \texttt{DEN REG}.
	\begin{figure}
		\centering
		\begin{tikzpicture}
			\begin{feynman}
				\tikzfeynmanset{every blob/.style={fill=gray, thick, draw=black}}
				\node at (0, 0) {$-q_{\mu}$};
				\draw[line width=1.1pt] (1,1.4) arc[start angle=110, end angle=250, x radius=0.5, y radius=1.55];
				\draw[line width=1.1pt] (5,1.4) arc[start angle=70, end angle=-70, x radius=0.5, y radius=1.55];
				\draw[line width=1.1pt] (7,1.4) arc[start angle=110, end angle=250, x radius=0.5, y radius=1.55];
				\draw[line width=1.1pt] (8.5,1.4) arc[start angle=70, end angle=-70, x radius=0.5, y radius=1.55];
				\node at (9,1.3) {$-1$};
				\draw[line width=1.1pt] (10.6,1.4) arc[start angle=110, end angle=250, x radius=0.5, y radius=1.55];
				\draw[line width=1.1pt] (11.5,1.4) arc[start angle=70, end angle=-70, x radius=0.5, y radius=1.55];
				\node at (12,1.3) {$-1$};
				\node at (1, 0) {$\mu$};
				\vertex (v1) at (1.2, 0);
				\vertex (v2) at (3, 0);
				\vertex (v3) at (3.5, 0);
				\vertex (vh4) at (3.5,1.5);
				\vertex (vh5) at (3.5,0.5);
				\vertex (vb6) at (3.5,-0.5);
				\vertex (vb7) at (3.5,-1.5);
				\node[blob, minimum size=1cm] at (v3) {};
				\node at (6.1,0) {$=\ e$};
				\node at (10,0) {$e$};
				\vertex (v8) at (7.5,0);
				\vertex (vh42) at (7.5,1.5);
				\vertex (vh52) at (7.5,0.5);
				\vertex (vb62) at (7.5,-0.5);
				\vertex (vb72) at (7.5,-1.5);
				\node[blob, minimum size=1cm] at (v8) {};
				\node at (9.4,0) {$-$};
				\vertex (v83) at (11,0);
				\vertex (vh43) at (11,1.5);
				\vertex (vh53) at (11,0.5);
				\vertex (vb63) at (11,-0.5);
				\vertex (vb73) at (11,-1.5);
				\node[blob, minimum size=1cm] at (v83) {};
				\diagram* {
					(v1) -- [boson, decoration={snake}, momentum=$q$] (v2),
					(vh5) -- [fermion, edge label'= $p+q$] (vh4),
					(vb7) -- [fermion, edge label'= $p$] (vb6),
					(vh52) -- [fermion, edge label'= $p+q$] (vh42),
					(vb72) -- [fermion, edge label'= $p+q$] (vb62),
					(vh53) -- [fermion, edge label'= $p$] (vh43),
					(vb73) -- [fermion, edge label'= $p$] (vb63),
				};
			\end{feynman}
		\end{tikzpicture}
		\caption{\texttt{WTI} diagram}
		\label{fig:WTI_diagram86486}
	\end{figure}
	\subsection{Setup and Perturbative Expansions}
	Gauge invariance implies an exact relation between the full (all-order) fermion-photon vertex $\Gamma^\mu$ and the inverse fermion propagator $S_F^{-1}$, as represented pictorially in Fig.~\ref{fig:WTI_diagram86486}, known as the \texttt{WTI}, reads:
	\begin{equation}\label{WTI2juin}
		-q_\mu\Gamma^\mu(p^\prime,p)=eS_F^{-1}(p^\prime)-eS_F^{-1}(p),\quad p^\prime=p+q.
	\end{equation}
	To \texttt{NLO} in perturbation theory, the fermion-photon vertex takes the form
	\begin{equation}\label{gamma2juin}
		\Gamma^\mu(p^\prime,p)=ie\gamma^\mu+ie\Gamma^\mu_{\text{1-Loop}}(p^\prime,p),
	\end{equation}
	and the fermion propagator $S_F(p)$, at \texttt{NLO} is obtained as follows:
	\begin{align}
		S_F(p)&=\frac{i(\cancel{p}+m)}{p^2-m^2+i\epsilon}+\frac{i(\cancel{p}+m)}{p^2+m^2+i\epsilon}\left(-i\sum_{\text{1-Loop}}(\cancel{{p}})\right)\frac{i(\cancel{{p}}+m)}{p^2+m^2+i\epsilon}\nonumber\\
		&=\frac{i}{\cancel{{p}}-m-\sum\limits_{\text{1-Loop}}(\cancel{{p}})},\label{prop2juin}
	\end{align}
	where, in the second line, we have summed the geometric series. Substituting the expansions \eqref{gamma2juin} and \eqref{prop2juin} into Eq.~\eqref{WTI2juin} at \texttt{NLO} yields
	\begin{equation}\label{beat56}
		-q_\mu \Gamma^\mu_{\text{1-Loop}}(p^\prime,p)
		= \sum\limits_{\text{1-Loop}}(\cancel{p}^\prime)-\sum\limits_{\text{1-Loop}}(\cancel{p}),
	\end{equation}
	where $\Gamma^\mu_{\text{1-Loop}}$ and $\sum\limits_{\text{1-Loop}}$ denote the one-loop vertex and self-energy corrections, respectively.
	
	Using the result from the \texttt{DEN REG} scheme, Eq.~\eqref{resultDENREG_VeRt}, the left-hand side of Eq.~\eqref{beat56} can be computed as
	\begin{equation}
		-q_\mu\Gamma^\mu_{\text{1-Loop}}(p^\prime,p)=-\cancel{q}F_1(q^2)-\frac{i}{2m}F_2(q^2)\left(q_\mu \sigma^{\mu\nu}q_\nu\right)=-\cancel{q}F_1(q^2),
	\end{equation}
	because $q_\mu \sigma^{\mu\nu}q_\nu=0$. Consequently, Eq.~\eqref{beat56} can be rewritten as
	\begin{equation}
		-\cancel{q}F_1(q^2)=\sum\limits_{\text{1-Loop}}(\cancel{p}+\cancel{q})-\sum\limits_{\text{1-Loop}}(\cancel{p}).
	\end{equation}
	Evaluating this expression on-shell, $p^2 = {p^\prime}^2 = m^2$, which corresponds to the limit $q^2 \to 0$, we obtain
	\begin{equation}
		\lim_{q^2\to 0}F_1(q^2)=\lim_{q^2\to 0}\left[-\frac{\sum\limits_{\text{1-Loop}}(\cancel{p}+\cancel{q})-\sum\limits_{\text{1-Loop}}(\cancel{p})}{\cancel{q}}\right].
	\end{equation}
	Hence, we obtain
	\begin{equation}\label{WTIshellneo}
		F_1(0) = -\frac{\partial }{\partial \cancel{p}}\Bigg(\sum\limits_{\text{1-Loop}} (\cancel{p})\Bigg) \Bigg|_{\cancel{p}=m}.
	\end{equation}
	Eq.~\eqref{WTIshellneo} represents the on-shell \texttt{WTI} at one loop, relating the Dirac form factor at zero momentum transfer, $F_1(0)$, to the momentum dependence of the electron self-energy. The derivative with respect to $\cancel{p}$ is evaluated on shell at $\cancel{p}=m$ and enforces the consistency between vertex and self-energy corrections required by gauge invariance.
	\subsection{One Loop Verification}
	We explicitly verify the relation at the one loop level, demonstrating its consistency with the \texttt{WTI} under on-shell conditions. In particular, we verify Eq.~\eqref{WTIshellneo}:
	\begin{equation}\label{key5lolo}
		F_1(0)=\frac{2e^2}{(4\pi)^2}\int_0^1\int_0^1\int_0^1 dxdydz\delta(x+y+z-1)\left[\mathbb{E}-2\ln(1-z)-\frac{2z}{(1-z)^2}\right]+\mathcal{O}(\epsilon),
	\end{equation}
	where
	\begin{equation}
		\mathbb{E}:=\frac{1}{\epsilon}-\gamma_E+\ln(4\pi)+\ln\left(\frac{\mu^2}{m^2}\right).
	\end{equation}
	We make use of the following integrals:
	\begin{equation}
		\int_0^1\int_0^1\int_0^1dxdydz\delta(x+y+z-1)=\frac{1}{2}\quad\text{and}\quad\int_0^1\int_0^1dxdy\delta(x+y+z-1)=1-z.
	\end{equation}
	Eq.~\eqref{key5lolo} becomes:
	\begin{equation}
		F_1(0)=\frac{2e^2}{(4\pi)^2}\left[\frac{1}{2}\mathbb{E}-2\int_0^1dz\ln(1-z)+2\int_0^1dzz\ln(1-z)-2\int_{0}^{1}dz\frac{z}{1-z}\right].
	\end{equation}
	The evaluation of the integrals relies on the standard series expansion of the logarithm. We expand $\ln(1-z)$ as a power series and integrate term by term, which yields a simple telescopic sum:
	\begin{equation}\label{4dfoinr6}
		\int_0^1dz\ln(1-z)=\sum_{n=1}^{\infty}\frac{-1}{n(n+1)}=-1\quad\text{and}\quad \int_0^1dzz\ln(1-z)=\sum_{n=1}^{\infty}\frac{-1}{n(n+2)}=-\frac{3}{4}.
	\end{equation}
	Thus, we find
	\begin{equation}\label{avantLASWTI}
		F_1(0)=\frac{e^2}{(4\pi)^2}\left[\mathbb{E}+1-4\int_{0}^{1}dz\frac{z}{1-z}\right].
	\end{equation}
	We now compute the right-hand side of Eq.~\eqref{WTIshellneo}. Let us denote
	\begin{equation}
		\mathbb{K}:=\frac{1}{\epsilon}-\gamma_E+\ln(4\pi)+\ln(\mu^2).
	\end{equation}
	Then, using our one-loop electron self-energy, we have:
	\begin{align*}
		-\frac{\partial }{\partial \cancel{p}}\Bigg(\sum\limits_{\text{1-Loop}} (\cancel{p})\Bigg) \Bigg|_{\cancel{p}=m} &=\frac{-2e^2}{(4\pi)^2}\int_{0}^{1}dx\frac{\partial}{\partial \cancel{p}}\bigg[\big(-x\cancel{p}+2m\big)\big(\mathbb{K}-\ln[(1-x)m^2-(1-x)xp^2]\big)\bigg]\Bigg|_{\cancel{p}=m}\nonumber\\
		&=\frac{2e^2}{(4\pi)^2}\left[\frac{1}{2}\mathbb{K}-\frac{1}{2}\ln(m^2)-2\int_0^1dxx\ln(1-x)-1-2\int_0^1dx\frac{x}{1-x}\right].
	\end{align*}
	Accounting for the second integral in Eq.~\eqref{4dfoinr6} and defining $\mathbb{E}:=\mathbb{K}-\ln(m^2)$, we obtain: 
	\begin{equation}\label{lasxprPART}
		-\frac{\partial }{\partial \cancel{p}}\Bigg(\sum\limits_{\text{1-Loop}} (\cancel{p})\Bigg) \Bigg|_{\cancel{p}=m}=\frac{e^2}{(4\pi)^2}\left[\mathbb{E}+1-4\int_0^1dx\frac{x}{1-x}\right]. 
	\end{equation}
	We thus see that Eqs.~\eqref{avantLASWTI} and \eqref{lasxprPART} are similar, completing the verification of the \texttt{WTI}. \hfill$\Box$
	
	\section{Conclusions}\label{cclDNWTI}
	We began by introducing the correspondence between the commonly used regularization scheme \texttt{DIM REG} and the recently proposed regularization scheme \texttt{DEN REG}. This correspondence was shown to be self-consistent and allowed us to determine the deterministic coefficient functions required for our context. We applied \texttt{DEN REG} to the one loop electron self-energy and vertex correction in \texttt{QED}. We explicitly verified that the \texttt{WTI} is satisfied, demonstrating that gauge symmetry is preserved within this regularization scheme. A particular advantage of \texttt{DEN REG} here is that it allows the numerator algebra to be performed directly in four dimensions, significantly simplifying the evaluation of one loop vertex corrections.
	
	Throughout the derivation, the principles of \texttt{DEN REG} were maintained, except for the determination of the coefficient functions. These were fixed by establishing a correspondence between \texttt{DIM REG} and \texttt{DEN REG}, leading to a deterministic prescription for most coefficients. An exception is the coefficient $g_{(3,1)}(\epsilon)$, which does not approach unity in the limit $\epsilon \to 0$. The investigation of such nontrivial coefficient functions is left for future work.
	
	From a physical standpoint, our verification of the \texttt{WTI} was carried out under on-shell conditions. An important extension of this analysis would be to test the \texttt{WTI} off shell, which could further clarify the scope and robustness of \texttt{DEN REG}. Overall, our results support \texttt{DEN REG} as a reliable and systematic alternative regularization scheme for symmetry-preserving perturbative calculations in \texttt{QFT}, with potential applications to more complex processes and higher-loop orders.
	\section*{Acknowledgement}
	The author would like to thank W. A. Horowitz for helpful discussions and for providing valuable feedback and comments on the notes I compiled while performing these calculations.
	
	\appendix
	\section{Some Important Integrals in \texttt{DEN REG}}\label{appendix_1}
	We summarize several integrals that frequently appear throughout the calculations in \texttt{DEN REG}:
	\begin{align}
		\int_0^\infty dk\frac{k^3}{(k^2+\Delta^2)^{2+\epsilon}}&=\frac{1}{2\epsilon(1+\epsilon)\Delta^{2\epsilon}},\label{integral_1}\\
		\int_0^\infty dk\frac{k^5}{(k^2+\Delta^2)^{2+\epsilon}}&=\frac{-1}{(1-\epsilon)\epsilon(1+\epsilon)\Delta^{2(\epsilon-1)}},\label{integral_2}\\
		\int_{0}^{\infty}dk\frac{k^3}{(k^2+\Delta^2)^{3+\epsilon}}&=\frac{1}{2(1+\epsilon)(2+\epsilon)\Delta^{2(\epsilon+1)}},\label{integral_3}\\
		\int_{0}^{\infty}dk\frac{k^5}{(k^2+\Delta^2)^{3+\epsilon}}&=\frac{1}{\epsilon(1+\epsilon)(2+\epsilon)\Delta^{2\epsilon}}\label{integral_4},
	\end{align}
	where $\Delta^2=\Delta^2\left(m^2;\{p_i^2\};\{x_j\}\right)$, with $m^2,p_i^2\in\mathbb{R}$ and ${x_j}$ denoting the Feynman parameters, $i,j\in\mathbb{Z}^{+}$. The integrals in Eqs.~\eqref{integral_1}--\eqref{integral_4} are derived using standard Beta function formulas and properties of the Gamma function.
	\section{Evaluation of $\mathcal{N}^\mu$ in $4$ Dimensions}\label{appendix_2}
		The aim of this appendix is to re-express the numerator in terms of $\ell$, using the \(\gamma\)-matrix algebra in four dimensions. Starting with Eq.~\eqref{num_VERTEX_Calc}, we have
		\begin{equation}\label{numFR}
			\mathcal{N}^\mu=\gamma^\nu(\cancel{p}^\prime+\cancel{k})\gamma^\mu(\cancel{p}+\cancel{k})\gamma_\nu+\gamma^\nu(\cancel{p}^\prime+\cancel{k})\gamma^\mu m\gamma_\nu+\gamma^\nu m\gamma^\mu(\cancel{p}+\cancel{k})\gamma_\nu+m^2\gamma^\nu\gamma^\mu\gamma_\nu.
		\end{equation}
		We employ the Dirac algebra in four dimensions to compute each term, we have the following:
		\begin{equation}
			\begin{aligned}
				\gamma^\nu(\cancel{p}^\prime+\cancel{k})\gamma^\mu(\cancel{p}+\cancel{k})\gamma_\nu &=-2(\cancel{p}+\cancel{k})\gamma^\mu(\cancel{p}^\prime+\cancel{k}),\\
				m\gamma^\nu(\cancel{p}^\prime+\cancel{k})\gamma^\mu \gamma_\nu&=4m(p^\prime+k),\\
				m\gamma^\nu\gamma^\mu(\cancel{p}+\cancel{k})\gamma_\nu&=4m(p+k),\\
				m^2\gamma^\nu\gamma^\mu\gamma_\nu&=-2m^2\gamma^\mu.
			\end{aligned}
		\end{equation}
		Then, Eq.~\eqref{numFR} become
		\begin{equation*}
			\mathcal{N}^\mu=-2m^2\gamma^\mu+4m(p^\prime+p+2k)^\mu-2(\cancel{p}+\cancel{k})\gamma^\mu(\cancel{p}^\prime+\cancel{k}).
		\end{equation*}
	We re-express this numerator in terms of $\ell:=k+xp+yp^\prime$ and we neglect the linear terms, we find
	\begin{equation*}
		\mathcal{N}^\mu=-2m^2\gamma^\mu+4m(p^\prime+p-2xp-2yp^\prime))^\mu-2\cancel{\ell}\gamma^\mu\cancel{\ell}-2(\cancel{p}-x\cancel{p}-y\cancel{p}^\prime)\gamma^\mu(\cancel{p}^\prime-x\cancel{p}-y\cancel{p}^\prime).
	\end{equation*}
	Using $x+y+z=1$ and $q=p^\prime-p$, the following manipulation are useful:
	\begin{equation*}
		2xp+2yp^\prime=(x+y)(p+p^\prime)+(x-y)(p-p^\prime),
	\end{equation*}
	which implies that
	\begin{align}
		p+p^\prime-2xp-2yp^\prime&=p+p^\prime-(x+y)(p+p^\prime)-(x-y)(p-p^\prime)\nonumber\\
		&=z(p+p^\prime)+(x-y)q.
	\end{align}
	And
	\begin{equation}
		\begin{aligned}
			p-xp-yp^\prime&=zp^\prime-(1-x)q\\
			p^\prime-xp-yp^\prime&=zp+(1-y)q.
		\end{aligned}
	\end{equation}
	Then
	\begin{equation}
		\mathcal{N}^\mu=-2m^2\gamma^\mu+4mz(p+p^\prime)^\mu+4m(x-y)q^\mu-2\cancel{\ell}\gamma^\mu\cancel{\ell}-2\mathcal{N}_1^\mu,
	\end{equation}
	where
	\begin{equation*}
		\mathcal{N}_1^\mu:=(z\cancel{p}^\prime+(x-1)\cancel{q})\gamma^\mu(z\cancel{p}+(1-y)\cancel{q}).
	\end{equation*}
	We know that the two Dirac spinors $\bar{u}(p^\prime)$ and $u(p)$ satisfy the following relations \cite{peskin, schwartz}:
	\begin{equation*}
		\bar{u}(p^\prime)\cancel{p}^\prime=\bar{u}(p^\prime)m\quad\text{and}\quad\cancel{p}u(p)=mu(p).
	\end{equation*}
	More generally in $\bar{u}(p^\prime)\Gamma^{\mu}(p^\prime,p)u(p)$, we have:
	\begin{equation}
		A\cancel{p}=Am \quad\text{and}\quad\cancel{p}^\prime B=mB \quad\text{for some $A$ or $B$}.
	\end{equation}
	Consequently
	\begin{align}
		\mathcal{N}_1^\mu&=(z\cancel{p}^\prime+(x-1)\cancel{q})\gamma^\mu(z\cancel{p}+(1-y)\cancel{q})\nonumber\\
		&=(zm+(x-1)\cancel{q})\gamma^\mu(zm+(1-y)\cancel{q}))\nonumber\\
		&=z^2m^2\gamma^\mu-(1-x)(1-y)\cancel{q}\gamma^\mu\cancel{q}+zm([\gamma^\mu,\cancel{q}]+x\cancel{q}\gamma^\mu-y\gamma^\mu\cancel{q}).
	\end{align}
	We use the following trick:
	\begin{equation}
		x\cancel{q}\gamma^\mu-y\gamma^\mu\cancel{q}=\frac{1}{2}(x-y)\{\gamma^\mu,\cancel{q}\}-\frac{1}{2}(x+y)[\gamma^\mu,\cancel{q}],
	\end{equation}
	which implies that
	\begin{align}
		\mathcal{N}_1^\mu&=z^2m^2\gamma^\mu-(1-x)(1-y)\cancel{q}\gamma^\mu\cancel{q}+zm\left([\gamma^\mu,\cancel{q}]+\frac{1}{2}(x-y)\{\gamma^\mu,\cancel{q}\}-\frac{1}{2}(x+y)[\gamma^\mu,\cancel{q}]\right)\nonumber\\
		&=z^2m^2\gamma^\mu-(1-x)(1-y)\cancel{q}\gamma^\mu\cancel{q}+\frac{1}{2}zm(2-x-y)[\gamma^\mu,\cancel{q}]+\frac{1}{2}zm(x-y)\{\gamma^\mu,\cancel{q}\}.\label{num1d}
	\end{align}
	Define the following formulas:
	\begin{equation}
		[\gamma^\mu, \gamma^\nu]\equiv-2i\sigma^{\mu\nu}\quad\text{and}\quad \{\gamma^\mu,\gamma^\nu\}\equiv2\eta^{\mu\nu}.
	\end{equation}
	For these reasons:
	\begin{equation}
		\frac{1}{2}[\gamma^\mu,\cancel{q}]=\frac{1}{2}(\gamma^\mu\cancel{q}-\cancel{q}\gamma^\mu)=\frac{1}{2}[\gamma^\mu,\gamma^\nu]q_\nu=-i\sigma^{\mu\nu}q_\nu,
	\end{equation}
	and
	\begin{equation}
		\frac{1}{2}\{\gamma^\mu,\cancel{q}\}=\frac{1}{2}(\gamma^\mu\cancel{q}+\cancel{q}\gamma^\mu)=\frac{1}{2}\{\gamma^\mu,\gamma^\nu\}q_\nu=q^\mu.
	\end{equation}
	One can easily calculate that $\cancel{q}\gamma^\mu\cancel{q}=-q^2\gamma^\mu$ because $\bar{u}(p^\prime)\cancel{q}u(p)=0$. Substituting the above results into Eq.~\eqref{num1d}, we obtain:
	\begin{align}
		\mathcal{N}_1^\mu&=z^2m^2\gamma^\mu+(1-x)(1-y)q^2\gamma^\mu-izm(2-x-y)\sigma^{\mu\nu}q_\nu+zm(x-y)q^\mu\nonumber\\
		&=z^2m^2\gamma^\mu+(z+xy)q^2\gamma^\mu-imz(1+z)\sigma^{\mu\nu}q_\nu+zm(x-y)q^\mu.
	\end{align}
	Putting together all numerator terms, we arrive at:
	\begin{equation*}
		\mathcal{N}^\mu=-2\cancel{\ell}\gamma^\mu\cancel{\ell}+4mz(p^\prime+p)^\mu-2(1+z^2)m^2\gamma^\mu-2(z+xy)q^2\gamma^\mu+2(x-y)(2-z)mq^\mu+2z(1+z)im\sigma^{\mu\nu}q_\nu.
	\end{equation*}
	We use the Gordon identity
	\begin{equation}
		(p^\prime+p)^\mu=2m\gamma^\mu-i\sigma^{\mu\nu}q_\mu.
	\end{equation}
	Then, we obtain
	\begin{equation*}
		\mathcal{N}^\mu=-2\cancel{\ell}\gamma^\mu\cancel{\ell}-2(1-4z+z^2)m^2\gamma^\mu-2(z+xy)q^2\gamma^\mu-2imz(1-z)\sigma^{\mu\nu}q_\nu+2(x-y)(2-z)mq^\mu.
	\end{equation*}
	We neglect the last term because it is antisymmetric under exchange of the variables $x\longleftrightarrow y$. Then
	\begin{equation}
		\mathcal{N}^\mu=-2\cancel{\ell}\gamma^\mu\cancel{\ell}-2(1-4z+z^2)m^2\gamma^\mu-2(z+xy)q^2\gamma^\mu-2imz(1-z)\sigma^{\mu\nu}q_\nu.
	\end{equation}
	According to \cite{peskin}, the following replacement is performed
	\begin{equation}
		\ell_\alpha \ell_\beta\longrightarrow\frac{1}{4}\eta_{\alpha\beta}\ell^2\quad\text{which implies}\quad\cancel{\ell}\gamma^\mu\cancel{\ell}\longrightarrow -\frac{1}{2}\ell^2\gamma^\mu.
	\end{equation}
	Our final expression for the numerator is given by
	\begin{equation}
		\mathcal{N}^\mu=[\ell^2-2\Delta^2+2z(2m^2-q^2)]\gamma^\mu-2imz(1-z)\sigma^{\mu\nu}q_\nu,
	\end{equation}
	where $\Delta^2:=(1-z)^2m^2-xyq^2$.
	
\end{document}